\input{aipcheck}

\documentclass[
    ,final            
  ]
  {aipproc}

\layoutstyle{8x11single}

\begin{document}

\title{Kinetic simulations of X-B and O-X-B mode conversion}


\author{A. V. Arefiev}{
  address={Institute for Fusion Studies, The University of Texas, Austin, Texas 78712, USA}
}

\author{E. J. Du Toit}{
  address={York Plasma Institute, Department of Physics, University of York , York, U.K.}
}

\author{A. K\"ohn}{
  address={IGVP, University of Stuttgart, Stuttgart, Germany}
  ,altaddress={Max Planck Institute for Plasma Physics, Garching, Germany} 
}

\author{E. Holzhauer}{
	  address={IGVP, University of Stuttgart, Stuttgart, Germany}
}

\author{V. F. Shevchenko}{
	  address={EURATOM/CCFE Fusion Association, Culham Science Centre, Abingdon, U.K.}
}

\author{R. G. L. Vann}{
  address={York Plasma Institute, Department of Physics, University of York , York, U.K.}
}

\begin{abstract}
We have performed fully-kinetic simulations of X-B and O-X-B mode conversion in one and two dimensional setups using the PIC code EPOCH. We have recovered the linear dispersion relation for electron Bernstein waves by employing relatively low amplitude incoming waves. The setups presented here can be used to study non-linear regimes of X-B and O-X-B mode conversion.
\end{abstract}

\maketitle


\section{Introduction}

High-performance spherical tokamaks are usually overdense (typically $\omega_{pe} / \omega_{ce} \sim 4$ in the core) and so regular electron cyclotron emission is blocked. However, (electrostatic) electron Bernstein waves, generated at harmonics of the local cyclotron frequency (and its harmonics) in the core may be observed outside the plasma via a mode conversion process to an electromagnetic mode~\cite{Laqua2007}. Understanding the details of this mode conversion process is important in tokamaks with over-dense plasmas both for the interpretation of microwave diagnostic data and to assess the feasibility of EBW heating and/or current drive~\cite{Urban2011}.


\section{Kinetic simulations using a particle-in-cell code}

Electron kinetics play an important role in the excitation and propagation of the Electron Bernstein waves (EBW). The linear regime has been studied extensively and is well understood. On the other hand, the EBW physics in the nonlinear regime, which is of interest in the context of heating and current drive, still presents a challenge~\cite{Xiang2011,Asgarian2014}. Analytical description of the nonlinear regime is not generally feasible and therefore numerical simulations are required.

A self-consistent description of electron kinetics and wave propagation must resolve spatial and temporal scales associated with electron gyro-motion in order to correctly recover the physics relevant to EBWs. One suitable approach is the particle-in-cell (PIC) framework that uses macro-particles to simulate electrons and ions kinetically. In the present work we use an MPI parallelized, explicit, second-order, relativistic code EPOCH~\cite{EPOCH} based on PSC~\cite{PSC}.

\section{Setups for studying EBW excitation}

\begin{figure} \label{Fig1}
  \includegraphics[height=.22\textheight]{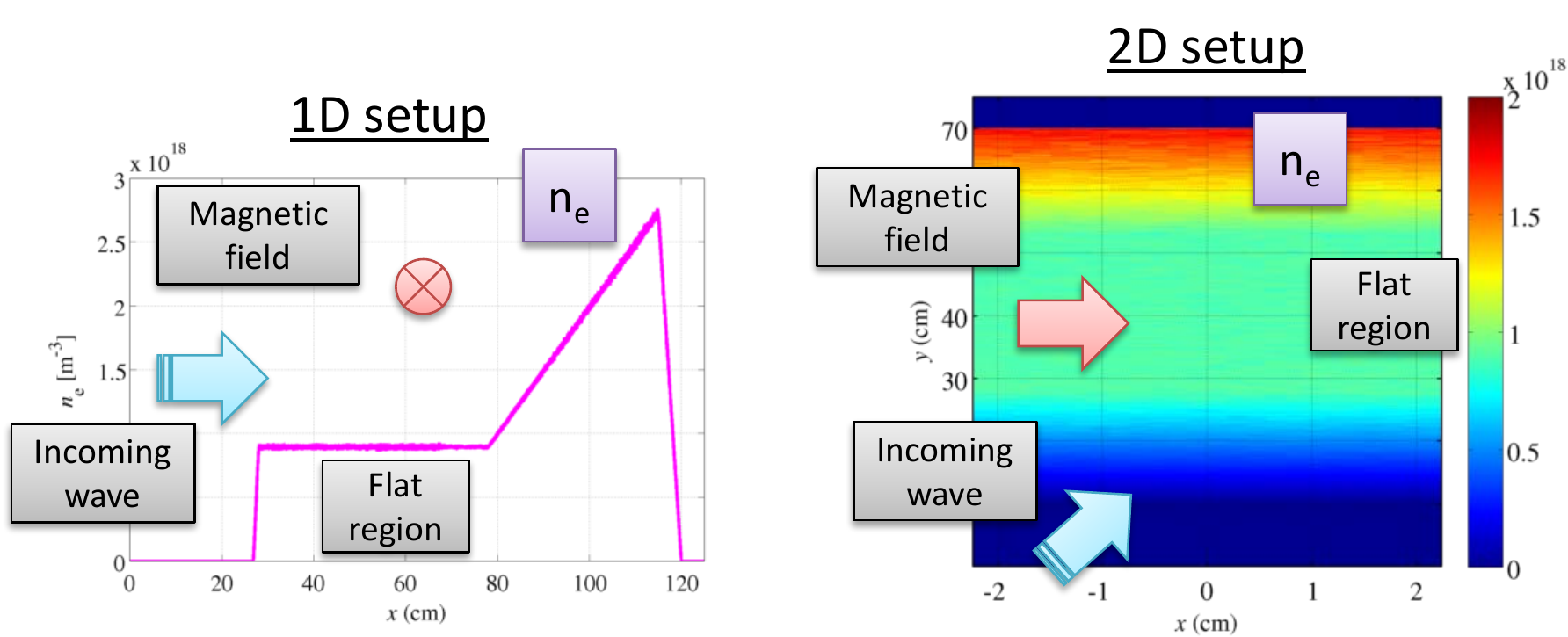}
  \caption{One and two-dimensional setups for PIC simulations of the X-B and O-X-B mode conversion.}
\end{figure}

We describe 1-D and 2-D setups that can be used to simulate EBW excitation using EPOCH. A suitable one-dimensional setup is shown in the left panel of Fig.~\ref{Fig1}. There are vacuum gaps between the plasma and the computational domain boundaries to reduce numerical boundary effects. A uniform magnetic field directed along one of the perpendicular coordinates ($y$ or $z$) is initialized throughout the domain. A plane electromagnetic wave is injected from the left side of the domain.

The plasma density profile has three distinct regions: 1) a sharp density gradient on the left; 2) an extended flat density region in the middle; 3) a high density region on the right. The sharp density gradient (first region) can be used to simulate the X-B mode conversion, achieved due to X-mode tunneling past the low-density cut-off and subsequent excitation of an EBW in the vicinity of the upper-hybrid resonance (UHR) layer. The extended flat density region (second region) is introduced to aid EBW identification. Its density should exceed the density corresponding to the UHR. Finally, the high density region that follows the flat region on the right is designed to introduce a high-density cut-off for the X-mode. If the density in the flat region is already sufficiently high, then the third region is not needed (which is the case in the 1D simulation of the X-B mode conversion described below). A density down-ramp that connects the high density region to the vacuum is recommended instead of a sharp discontinuity to reduce the numerical noise there. We use open boundary conditions for the fields and particles at both boundaries.  

A two-dimensional setup suitable for simulating EBW excitation is shown in the right panel of Fig.~\ref{Fig1}. There are obvious parallels between this setup and the one-dimensional setup described above: a vertical cross-section of the two-dimensional domain looks similar to the one-dimensional setup. The key difference is that the 2D setup allows us to launch a wave at an angle to the magnetic field. A full 2D simulation where the incoming wave packet has a finite width is possible in principle. However, such a simulation can be computationally demanding, since the box has to be sufficiently wide to accommodate the path of a wave packet injected under an angle. A ``reduced'' 2D setup is thus needed in order to be able to perform multiple runs that are necessary for parameter scans.  Such a reduced setup is achieved by using periodic boundary conditions (for particles and fields) on the left and right side of the domain. The injected wave is a plane wave with given wave-vector components parallel ($k_{\parallel} \equiv k_x$) and perpendicular ($k_{\perp} \equiv k_y$) to the magnetic field. The transverse size of the domain (along the $x$ axis) is set to $2 \pi / k_{\parallel}$ to account for the wave periodicity. Similarly to the 1D setup, we use open boundary conditions at the top and bottom boundaries of the simulation domain. The plasma density again has three regions, which can be convenient for simulating the O-X-B mode conversion. The O-X-B conversion in this setup is achieved by setting the density in the flat region to be above the density for the UHR and below the critical density. In this case, the extended flat density region is again designed to aid EBW identification. 

Ultimately, the following considerations should be kept in mind when choosing the density in the flat region in both setups: 1) if O and X modes are not desired in the flat region, then its density should be above the high density cutoff; 2) the wavelength of the EBW that decreases with density must be large enough compared to the grid size; 3) the amplitude of the numerical field fluctuations that increases with density for a fixed number of macro-particles per cell must be much less than the amplitude of the expected EBW signal.

\section{X-B conversion in linear regime}

\begin{figure} \label{Fig2}
  \includegraphics[height=.34\textheight]{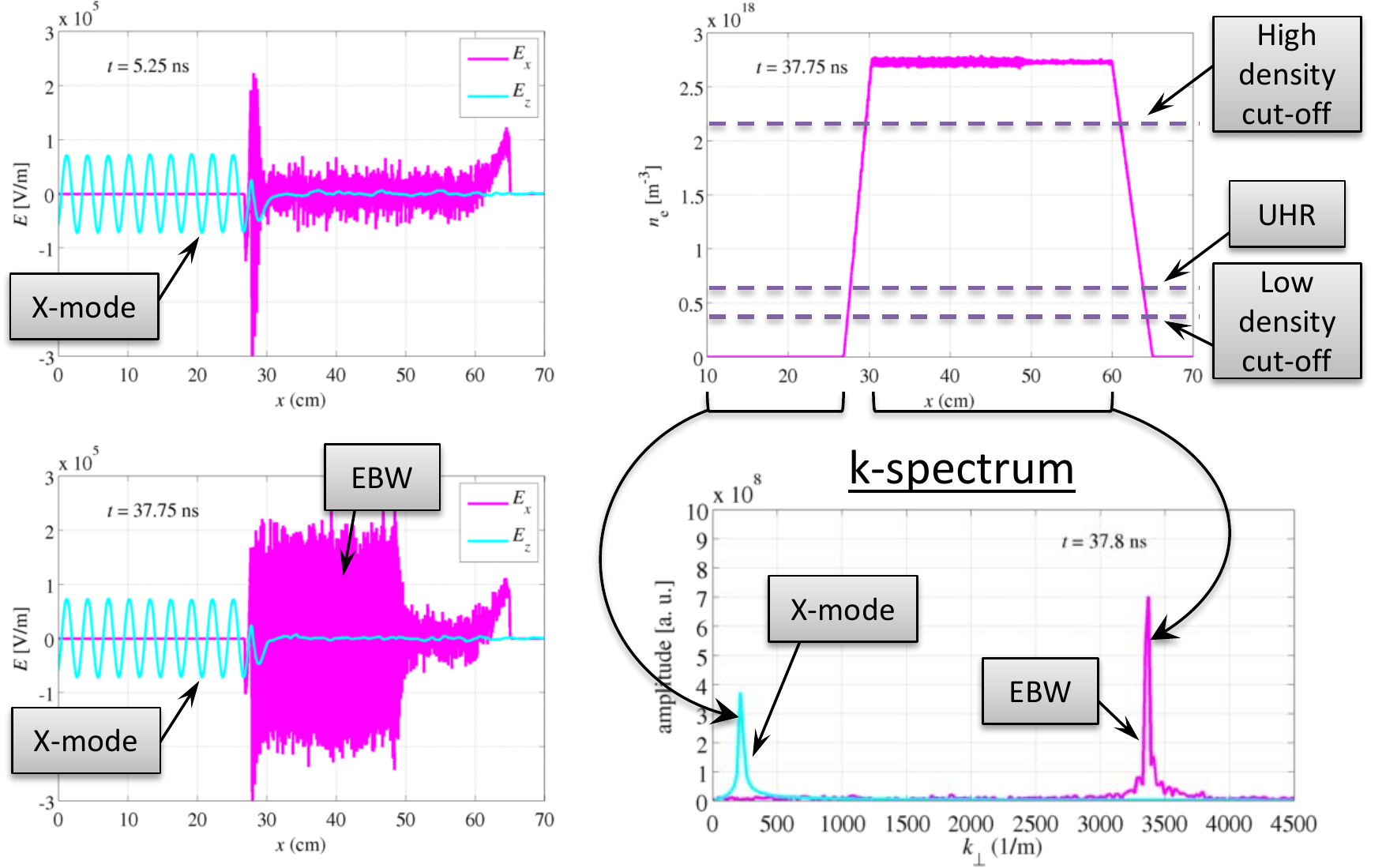}
  \caption{One-dimensional PIC simulation of the X-B mode conversion in a linear regime.}
\end{figure}

An example of the X-B mode conversion simulated in a 1D setup described above is shown in Fig.~\ref{Fig2}. A  confining uniform magnetic field of 0.25 T is directed along the $y$-axis. An incoming plane wave is an X-mode with electric field polarized along the $x$-axis. Its frequency is 10 GHz and its maximum amplitude is $10^5$ V/m. We ramp up the wave amplitude as a semi-Gaussian with a 1 ns width [we define a semi-Gaussian as $\exp( - (t-t_0)^2/w^2)$ for $t<t_0$, where $w$ is the width]. The electron density in the flat region is set at $n_e = 2.2 n_{\mbox{crit}}$, where $n_{\mbox{crit}}$ is the critical density. In this case, the low-density cut-off, the UHR, and the high-density cut-off are all below the density in the flat region (see the upper-right panel in Fig.~\ref{Fig2}). The density gradient at the left plasma boundary is $64 n_{\mbox{crit}}$m$^{-1}$. Such a steep density gradient is required for efficient EBW excitation. The electron temperature is set at $T_e = 1$ keV, while the ions are treated as immobile. There are 24000 macro-particles in each cell representing electrons. The cell size is $\Delta x = 7.8 \times 10^{-3}$ cm.

The left two panels in Fig.~\ref{Fig2} show snapshots of transverse and longitudinal components of the electric field. As expected, there is no X-mode tunneling into the flat density region. On the other hand, a short-scale longitudinal electric field is excited at the left density gradient and it then slowly propagates to the right. In order to identify the modes, we have performed a spatial Fourier transform of the electric field. The $k$-spectrum of the transverse electric field in the vacuum gap to the left of the plasma is shown with a cyan curve in the lower-right panel in Fig.~\ref{Fig2}, whereas the $k$-spectrum of the longitudinal field in the flat density region is shown with a magenta curve. The two narrow peaks match the linear dispersion relation for the X-mode and for the EBW. 

\section{O-X-B conversion in linear regime}

\begin{figure} \label{Fig3}
  \includegraphics[height=.44\textheight]{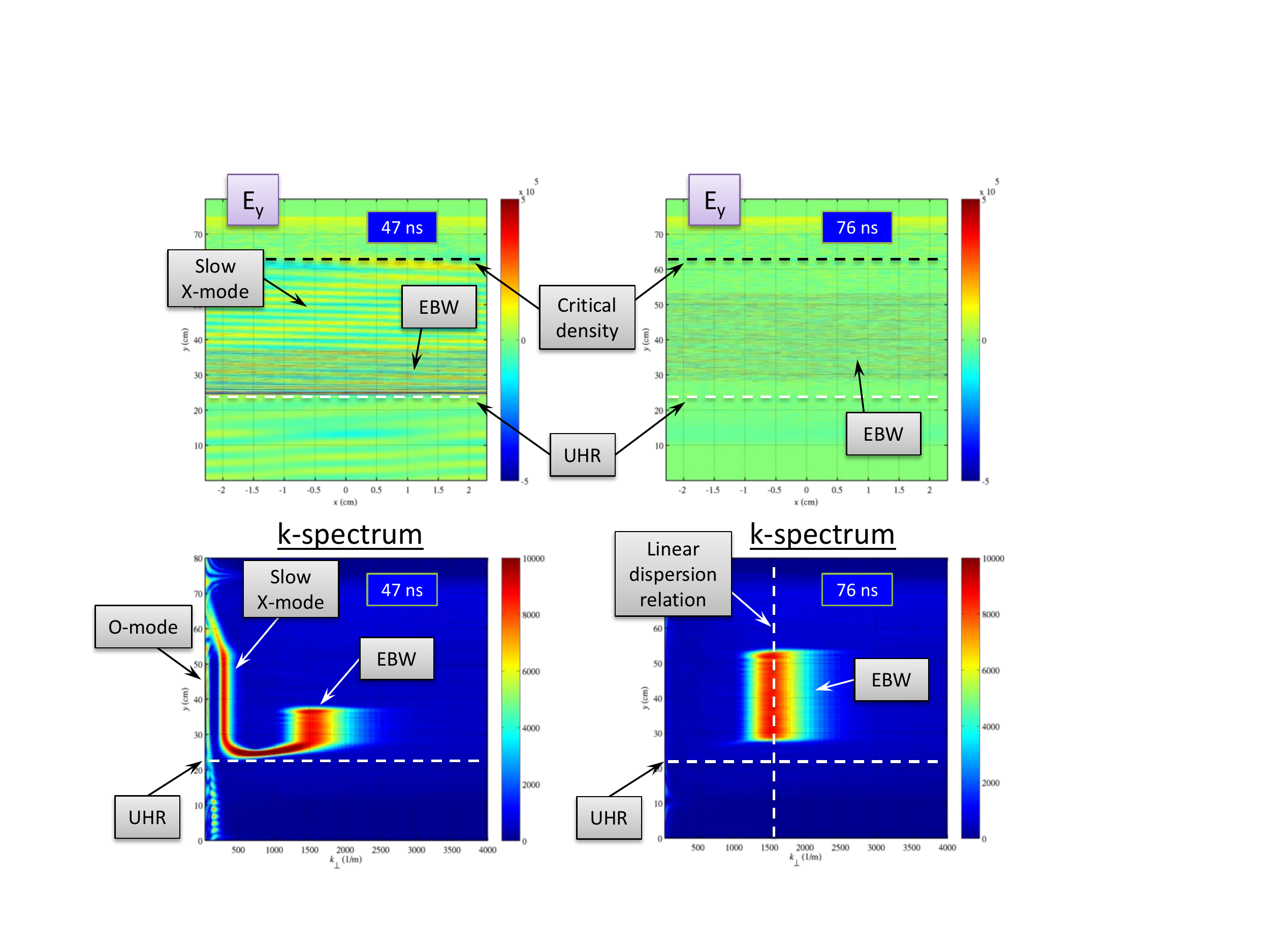}
  \caption{Two-dimensional PIC simulation of the O-X-B mode conversion in a linear regime.}
\end{figure}

An example of the O-X-B mode conversion simulated in a 2D setup described above is shown in Fig.~\ref{Fig3}. A confining uniform magnetic field of 0.25 T is directed along the $x$-axis. A plane wave with a maximum amplitude of $5 \times 10^4$ V/m is launched from the lower boundary at 40\textdegree (the angle between the wave front and the magnetic field). Its frequency is 10 GHz and its electric field is polarized in the plane of the simulation. We again ramp up the wave amplitude as a semi-Gaussian with a 1 ns width. The amplitude then remains constant for 50 ns after which it is ramped down to zero over 1 ns. The shape of the electron density profile is shown in the right panel of Fig.~\ref{Fig1}. The electron density in the flat region is set at $n_e = 0.72 n_{\mbox{crit}}$, where $n_{\mbox{crit}}$ is the critical density. In this setup, the density in the flat region is below the critical density and above the density corresponding to the UHR (see upper two panels of Fig.~\ref{Fig3}). The electron temperature is $T_e = 1$ keV, while the ions are treated as immobile. The electrons are represented by 800 macro-particles in each cell. There are 50 cells along the $x$-axis and 11200 cells along the $y$-axis ($\Delta y = 7.1 \times 10^{-3}$ cm). 

The upper two panels in Fig.~\ref{Fig3} show snapshots of $E_y$, whereas the lower two panels show the corresponding $k_y$-spectra as functions of $y$. The left panels clearly illustrate the O-X-B conversion, with all three modes (O-mode, slow X-mode, and EBW) being clearly visible in the $k$-spectrum of the flat density region. The right panels demonstrate that EBW indeed propagates towards the high density region without appreciable energy loss. The vertical dashed line in the lower-right panel is $k_y$ for EBW in the flat density region according to the linear dispersion relation. The simulated $k_y$ for the EBW agrees well with the linear dispersion relation.

\section{Summary}

We have proposed two setups that can be used in PIC simulations to study X-B and O-X-B mode conversion in one and two dimensions. Using these setups, we have performed PIC simulations of EBW excitation by waves with a relatively low amplitude ($10^5$ and $5 \times 10^4$ V/m). We have reproduced X-B and O-X-B mode conversion, with both the initial electromagnetic mode and the excited EBW matching the linear dispersion relations. The setups used in this work can now be employed to study EBW excitation in non-linear regimes.


\begin{theacknowledgments}
This work was funded in part by the US Department of Energy under grant DE-FG02-04ER54742, the University of York, the UK EPSRC under grant EP/G003955, and the European Communities under the contract of Association between EURATOM and CCFE. Simulations were performed using the EPOCH code (developed under UK EPSRC grants EP/G054940, EP/G055165 and EP/G056803) using HPC resources provided by the Texas Advanced Computing Center at The University of Texas and using the HELIOS supercomputer system at Computational Simulation Centre of International Fusion Energy Research Centre (IFERC-CSC), Aomori, Japan, under the Broader Approach collaboration between Euratom and Japan, implemented by Fusion for Energy and JAEA.
\end{theacknowledgments}



\bibliographystyle{aipproc}   




\end{document}